# Effect of polariton-polariton interactions on the excitation spectrum of a non-equilibrium condensate in a periodic potential

D. N. Krizhanovskii[1], E. A. Cerda[2], S. S. Gavrilov[3], D. Sarkar[1], K. Guda[1], R. Bradley[1], P. V. Santos[2], R. Hey[2], K. Biermann[2], M. Sich[1], F. Fras[1], M. S. Skolnick[1]

[1]Department of Physics and Astronomy, University of Sheffield, Sheffield S3 7RH, UK
[2] Paul-Drude-Institut fur Festkorperelektronik, Berlin, Germany
[3]Institute of Solid State Physics, Chernogolovka, Moscow region, Russia, 142432

**Polariton condensates are investigated in periodical potentials created by surface acoustic waves using both resonant and non-resonant optical excitation. Under resonant pumping condensates are formed due to polariton parametric scattering from the pump. In this case the single particle dispersion in the presence of the condensate shows a strong reduction of the energy gap arising from the acoustic modulation, indicating efficient screening of the surface acoustic wave potential by spatial modulation of the polariton density. The observed results are in good agreement with a model based on generalised Gross-Pitaveskii equations with account taken of the spatial dependence of the exciton energy landscape. In the case of incoherent, non-resonant pumping coexisting non-equilibrium condensates with s- and p- type wavefunctions are observed, which have different energies, symmetry and spatial coherence. The energy splitting between these condensate states is also reduced with respect to the gap of the one particle spectrum below threshold, but the screening effect is less pronounced than in the case of resonantly pumped system due to weaker modulation of the pump state**.

Microcavity polaritons are composite particles arising from strong exciton-photon coupling in semiconductor microresonators. Due to their excitonic component, polariton-polariton interactions are very strong, enabling the observation of stimulated polariton-polariton scattering[1,2], non-equilibrium polariton Bose-Einstein condensation (BEC)[3,4] superfluidity[5] and polariton solitons[6,7]. Polariton-polariton interactions also define the size of vortices formed in polariton condensates[8] and modify the excitation spectrum of the condensates, leading to a diffusive Goldstone mode[4,9].

Recently, investigation of exciton-polaritons in a periodic potential became a topic in its own right. Study of the spatial coherence of a non-equilibrium polariton condensates arising from optical parametric oscillation (OPO) and condensate transport using surface acoustic wave (SAW) have been reported[10]. Polariton condensation into p and d states under incoherent excitation has been observed in microcavities where weak 1D and 2D periodic potentials are created by metal patterning on the surface of microcavity[11,12]. Furthermore, the formation of bright polariton solitons[13] in spatially modulated microcavity wires has been reported[14]. As well as exciton-polaritons, transport of indirect excitons in semiconductor quantum wells using surface acoustic waves[15] and electrode conveyor belt were also demonstrated[16]. We note that investigation of equilibrium cold atom condensates in periodic lattices has also been very productive with observation of superfluid-to-Mott insulator transition[17], particle number squeezing and bright gap solitons.[18]

Without the periodic potential, polariton-polariton interactions have been shown to determine the single particle spectrum in the presence of a condensate, which can be a diffusive Goldstone mode[4,9] or a sound like Bogoliubov dispersion[19] depending on the excitation conditions. For the equilibrium self-interacting BEC in a periodic potential the interparticle interactions result in a linear sound-like dispersion at small wavevectors, but Brillouin zone structure in the excitation spectrum with opened gaps at higher momenta was shown to remain[20].

In this paper we study non-equilibrium interacting condensates in an external periodic potential induced by surface acoustic waves (SAWs), which coexist together with a large density in an external reservoir which is also subject to interactions with the potential and with the condensate. Polariton condensates using resonant and non-resonant excitation were addressed. In the case of resonant excitation into the lower polariton branch a polariton condensate (OPO signal) forms due to polariton-polariton pair scattering from the pump. Incoherently pumped condensates were formed under resonant excitation of the exciton-like lower polariton branch due to multiple polariton-polariton and polariton-phonon scattering.

Under resonant pumping, the condensate single particle excitation spectrum shows no observable gap due to spatial modulation, indicating efficient screening of the SAW potential by the modulated polariton density in the system. Further insight into the reasons behind the screening is obtained by measuring the intensity of the diffracted pump laser beam at different SAW amplitudes and optical densities. The spatial modulation of the highly populated pump state strongly affects the effective potential which is seen by single particle excitations. In the case of incoherent non-resonant pumping non-equilibrium coexisting condensates with s- and p- type wavefunctions are observed, which have different energies, symmetry and spatial coherence due to the non-equilibrium character of the polariton system. The energy splitting between these condensate states is also reduced with respect to the gap of the one particle spectrum below threshold, but the screening effects are less pronounced than in the case of the resonantly pumped system probably due to smaller modulation of the pump state under these excitation conditions.

**Experimental technique and samples**

The studies were carried out on GaAs-based microcavities grown by molecular beam epitaxy.[10] The condensates were generated in two distinct ways, namely, pumped by continuous wave (CW) lasers in resonance with the lower polariton branch and with the exciton reservoir, respectively, which lead to macroscopically occupied states (condensates) arising either from OPO[1,2] or incoherent polariton relaxation[21]. OPO condensates result from a parametric scattering process in which two pump polaritons scatter into signal and idler polaritons, fulfilling simultaneously energy and momentum conservation (Fig.1c). As a result, three coupled macroscopically occupied states (pump, signal and idler) are formed. It is important to note that the phase of the signal in the OPO forms spontaneously and is independent of that of the pump laser, which implies the breakdown of U(1) symmetry at the phase transition as in a standard BEC[9]. For excitation resonant with the exciton-like polariton states, by contrast, a non-equilibrium Bose-Einstein Condensate (BEC) forms through the incoherent relaxation of polaritons from a high energy exciton reservoir; this takes place mainly via polariton-exciton and polariton-polariton scattering[22]. Although the OPO signal and BEC are created in different ways, they exhibit very similar characteristic properties such as long spatial[10] and temporal coherence[23]. The spatial modulation of the polariton potential by acoustic fields was introduced by surface acoustic waves (SAWs) of wavelength $\lambda_{SAW}$ = 8 μm propagating along a non-piezoelectric [100] direction (the y-direction in Fig. 1a) [10]. The experiments were performed at 5 K on a region of the sample surface with nearly zero energy detuning between the excitonic and photonic modes at wavevector k = 0. Below we express the amplitude of the SAW potential *u* as the square root of the external nominal rf power applied to the inter-digitated transducer generating the SAW on the sample.

**Polariton Excitations in OPO condensates**

We first investigate OPO condensates resonantly excited by a pump laser (p) with in-plane wave vector $k_p = (k_{px}, k_{py}) = (1.4, 0)$ µm$^{-1}$ = $(1.8, 0) \times k_{SAW}$ (Fig. 1b). For low optical excitation densities well below the OPO threshold ($I_{OPO(th)} \sim 600$ W/cm$^2$) the polariton dispersion is obtained from the measurement of far-field photoluminescence (PL). Figure 2a shows the polariton PL intensity versus energy and momentum $k_y$ at applied amplitude of SAW u = 2.81 mW$^{0.5}$. There is a well-resolved energy gap of about $\Delta E_{gap}$ = 180 µeV observed at the edge of the first mini-Brillouin zone (MBZ), at $\pm k_{SAW}/2 = \pm \pi/\lambda_{SAW}$, indicating the formation of a polariton lattice in the presence of the SAW[24]. To a first order approximation the value of this gap is equal to half the peak-to-peak amplitude of the periodic potential $V_{SAW}$ induced by the acoustic wave. A further perturbation of the LP dispersion is observed at the edge of the second MBZ $\pm k_{SAW}$, where the band gap due to spatial modulation is significantly smaller.

For optical powers close to the condensation threshold polariton-polariton pair scattering from the pump is permitted into a wide range of states $0 \leq k \leq 2k_p$ obeying energy and momentum conservation rules. In planar un-modulated structures it was observed that above the threshold self-organisation occurs, leading to the formation of a macroscopically occupied signal condensate at $k \sim 0$ containing 500−1000 particles as well as a weaker idler at $k \sim 2k_p$[25]. In the case of the applied SAW potential, the OPO behaviour is different. The SAW induces a spatial modulation of both the signal and pump states, thus creating weaker pump polaritons with wave vectors $k = (1.4, \pm 1)k_{SAW}$ due to diffraction, as well as modulation of the signal. Emission from the diffracted polaritons has intensity ≤ 4% of the main pump mode at $k_p$.

Figures 2b through d display energy vs. $k_y$ images of the OPO signal emission recorded under increasing SAW powers for a fixed optical excitation power density of $I_L = 2I_{OPO}(th) = 1.2 \times 10^3$ W/cm$^2$ distributed over a 50 µm diameter Gaussian spot. Without the SAW, a condensate (Fig.2b) is formed at $k = 0$ and energy of 1.5345 eV, approximately 0.25 meV above the bottom of the lower polariton branch below threshold. The blue-shift arises from inter-particle polariton-polariton interactions[26,27]. As was observed before (Ref.[10]) at intermediate SAW amplitudes (u = 2.28 mW$^{0.5}$) the emission peaks not around $k_y = 0$ but at $k_y = \pm k_{SAW}/2$, corresponding to states at the boundary of the first MBZ(Fig.2a). It was suggested in ref 10 that such condensation proceeds by scattering between the main pump beam and its weaker diffracted replicas, which provides an additional gain channel[10]. The application of the SAW potential leads to spatial modulation of the extended condensate, where polariton particles concentrate mainly at the minima of the SAW potential. It was also revealed that at these intermediate SAW powers the $g^1$-spatial correlation function of the OPO signal is also spatially modulated and extends over 2-3 SAW periods[28]. The corresponding condensate emission with maxima at $k_y = \pm k_{SAW}/2$ suggests that the phase difference between the condensates in adjacent potential minima is equal to π, rather than zero[28]. Finally, at higher SAW amplitudes (u ≥ 2 mW$^{0.5}$) the momentum of the condensate is no longer well-defined and the emission pattern in momentum space spreads over the whole MBZ branch due to the stronger confinement. It is observed that at this SAW amplitude the spatial coherence of the OPO signal is reduced down to the size of a single potential well (~$\lambda_{SAW}/2$=4 μm)[10].

We now focus on the polariton emission at energies and momenta higher than that of the condensate emission. This emission has intensities 50-70 times smaller than that of the condensate in the first MBZ and represents the dispersion of a single particle excitation spectrum[29] in the OPO system in the presence of the condensate. It is seen from Figs.2(b-d) that no visible energy gap between the condensate and the higher energy emission is observed at $k_y = \pm k_{SAW}/2$. By superimposing the one particle polariton dispersion measured experimentally below the condensation threshold (thick white lines), on the condensate emission spectrum in momentum space it is seen that an energy gap (if one exists) due to spatial modulation is strongly reduced compared to that below threshold. For example, the experimental single particle excitation spectrum above threshold at k>$k_{SAW}/2$ is about 70 μeV below the white line dispersion measured below threshold in Fig.2d, indicating the gap in the excitation spectrum is smaller than ~80 μeV. Below we argue that the reduction of the gap arises from efficient screening of the SAW potential by the modulated polariton density: condensate single particle excitations interact with both the acoustic wave and the modulated polariton density, which act in opposite senses to one another, leading to a reduced effective polariton potential.

In the OPO regime, the contact polariton-polariton interaction leads to the energy renormalisation of the one particle excitation spectrum of the condensate [27, 29]. Given that the pump polariton density is an order of magnitude larger than that of the signal [23], the signal-pump interaction is likely to be dominant in the renormalisation of the OPO excitation spectrum. Below we provide a quantitative analysis of the mechanism responsible for the screening of the SAW potential in the OPO regime. Consider the case of spatially unmodulated polaritons. A single-mode model, describing the self-interacting pump mode polaritons excited with an external laser and taking into account the detuning of the external pump frequency with respect to that of the LP branch, predicts a bistable behavior of the pump polariton population as a function of the external laser power[27, 29]. In our case, the parameters of the system are such that the bistability threshold usually coincides with the OPO condensation threshold[30]. According to this model, at increased detuning of the pump frequency with respect to the LP branch, a stronger optical excitation power and hence a higher pump polariton density is required to bring the LP mode into resonance with the laser and to switch on the OPO. Once such a critical pump polariton density is created it becomes difficult to inject more pump polaritons into the system, since a higher polariton population blue shifts the polariton resonance away from the laser frequency. Thus, above the threshold the polariton density grows slowly and sublinearly with respect to the pump power[27,29]. Polariton modulation by the SAW results in a spatial modulation of the frequency detuning between the pump and the lower polariton branch resonance. Following the discussion above, a higher pump polariton population in the SAW minima than that in the SAW maxima is created at powers above the OPO threshold, leading to a spatial modulation of the pump polariton density in antiphase with the SAW potential. As a result, an effective periodic potential experienced by polaritons propagating in the presence of the SAW potential and the modulated pump density is reduced with respect to the SAW potential only.

Further insight into the behaviour of the polariton pump density and hence of the effective polariton potential as a function of the pump and SAW power can be obtained by studying the 1$^{st}$ order diffraction replicas of the pump beam. Their intensities are determined by the magnitude of the pump field inside the microcavity as well as by the spatial modulation of this field due to interaction with the SAW. Fig.3a shows the intensity of the diffracted pump beams versus external pump power $I_L$ at fixed SAW amplitude (u=2.8 mW$^{0.5}$). It first increases linearly with $I_L$ for pump power $I_L$ <300 W/cm$^2$ below the nonlinearity threshold, since the pump field inside the microcavity is a linear function of $I_L$ and the SAW amplitude. However, the bistability of the internal pump field described above leads to sublinear behaviour and saturation of the diffracted beam intensities above the OPO threshold at $I_L$>600 W/cm$^2$: indeed, the amplitude and the modulation depth of the pump density are expected to vary only very slightly at higher pump powers, whereas the phase of its modulation is also shifted by π with respect to the SAW modulation.

We also investigated the intensities of the diffraction replica at fixed pump $I_L$~800 W/cm$^2$ above OPO threshold versus SAW amplitude. It is observed to increase linearly with u at SAW amplitudes u<1.5 mW$^{0.5}$ as shown in Fig.3 b). Again, this result is consistent with our qualitative consideration: the higher the SAW amplitude, the stronger the modulation of the polariton density (which compensates the SAW potential), leading to higher intensity of the diffraction replica. At higher SAW amplitudes slight thermal broadening of the polariton states as well as an energy redshift of the polariton dispersion may lead to reduction of the modulation depth, which qualitatively explains the sublinear dependence of the intensity of the diffraction replicas at higher SAW amplitudes (Fig.3 b).

**Numerical modelling of OPO condensates in periodical potential**

Above we used a qualitative bistability model for the pump mode field as a function of pump power to explain the reduction of the effective potential due to the modulated polariton density in the presence of the SAWs.

In reality the OPO system is more complicated: the internal pump field depends not only on the cavity parameters and pump energy but also on the population of signal and idler states. Here we perform a simulation of the OPO system subject to a periodic potential using a semi-classical model based on many-mode 2D Gross-Pitaevskii equations. Such a model is well suited for a realistic description of the OPO regime in cavity-polariton systems. The effect of the SAW is taken into consideration by pre-setting an explicit sinusoidal distribution of the exciton frequency in the $y$ direction.

In the scalar ("spinless") approximation,

$$i\partial_t \psi_C(\mathbf{r},t) = \left(\omega_C^{(0)} - \frac{\hbar^2 \nabla^2}{2m_C} - i\gamma_C\right)\psi_C(\mathbf{r},t) + g\psi_X(\mathbf{r},t) + \alpha \mathcal{E}(\mathbf{r},t)e^{-i\omega_p t} \quad (1)$$

$$i\partial_t \psi_X(\mathbf{r},t) = g\psi_C(\mathbf{r},t) + \left[\omega_X^{(0)} + V_{\text{SAW}}\cos(\mathbf{k}_{\text{SAW}}\mathbf{r}) - i\gamma_X\right]\psi_X(\mathbf{r},t)$$
$$+ G\psi_X(\mathbf{r},t)^* \psi_X(\mathbf{r},t)\psi_X(\mathbf{r},t) \quad (2)$$

Here, $\psi_{C,X}$ are the amplitudes of the photon and exciton components of the intra-cavity field, $\mathcal{E}(\mathbf{r})$ is the amplitude of the pump wave with frequency $\omega_p$, $\alpha$ the response coefficient, $g$ the exciton-photon coupling strength (so that the Rabi splitting equals $2g$), and $G$ the exciton-exciton interaction constant. 2D photons and excitons are characterized by the $\mathbf{k}=0$ frequencies $\omega_{C,X}^{(0)}$ and decay rates $\gamma_{C,X}$; the exciton effective mass $m_X$ is much larger than that of cavity photon ($m_C$), so that the exciton free energy term ($-\hbar^2 \nabla^2/2m_X$) is neglected. The SAW periodical potential is characterized by its amplitude $V_{\text{SAW}}$ and wave vector $\mathbf{k}_{\text{SAW}}$.

In the simulations we assume $m_C = 3 \times 10^{-5} m_e$, $m_e$ being the free electron mass, $\gamma_C = 0.25$ meV and $\gamma_X = 0.1$ meV, $\omega_C^{(0)} = \omega_X^{(0)}$ (so that photon-exciton mismatch at $\mathbf{k}=0$ is zero), and Rabi splitting $2g = 6$ meV. The pump is a spatially uniform wave with in-plane wave number of $k_p = k_{px} = 1.4$ $\mu$m$^{-1}$ and uniform time dependence. $\hbar\omega_p$ is 0.25 meV larger than the unperturbed LP resonance at $k = k_p$. $G = 1$, so that $|\psi_X|^2$ has the dimension of energy. The effective potential has a period of $\lambda_{\text{SAW}} = 8$ $\mu$m and is applied in the $y$ direction, so that $\mathbf{k}_{\text{SAW}} = (0, 2\pi/\lambda_{\text{SAW}})$. It is characterized by the modulation amplitude $V_{\text{SAW}} = 0.3$ meV for the exciton component; since $\omega_C^{(0)} = \omega_X^{(0)}$, the polariton modulation potential appears 2.0 and 1.5 times smaller at $k = 0$ and $k = k_p$, respectively.

Figure 4a represents the PL spectrum under very weak excitation, revealing the SAW-modulated polariton dispersion. In this case the polariton emission arises mainly from zero-mean field fluctuations (Langevin noise), and we took very small spectral linewidths $\gamma_{C,X}$ for better representation. The spectrum has a well pronounced Brillouin zone structure with a band gap of approximately 0.15 meV. Fig. 4b shows the spectrum at a pump intensity near the threshold for parametric generation, where the momentum-space distribution of the signal becomes very narrow. It is seen that the dominant OPO signals appear in the $k_y = \pm k_{SAW}/2$ directions in agreement with the experimental observation [27]. We note that in real space the calculated signal intensity (not shown) exhibits maxima at the minima of the SAW potential, whereas the difference between the phases of signal wavefunction in adjacent potential minima is π. This is consistent with a condensation into s' states at the edge of the 1st MBZ (Fig.1b). Thus, in the case of resonant pumping the `symmetry' of the condensate appears to be imposed by the resonance blue-shifts. By contrast, incoherently pumped condensates (see Ref. [11] and the next Section) form in a p-state that shows intensity maxima at the maxima of the SAW potential.

The model above allows one to extract the steady-state spatial distribution of polariton density (taking into account the inter-mode polariton-polariton scattering) and, hence, to track the screening of the SAW potential due to the blue-shift of the polariton energy with increasing pump power. Fig. 5 represents the low-field polariton energy perturbation due to the SAW at $k = 0$, $\Delta E_{SAW}(y) = \frac{1}{2} V_{SAW} \cos(k_{SAW} y)$ (here a factor ½ accounts for the 50% exciton content in the polariton eigenstate), the blueshift due to exciton-exciton interactions, $\Delta E_{shift}(y) = \frac{1}{2} G |\psi_X(x = 0; y)|^2$, and the resultant $k = 0$ effective polariton energy landscape, $\Delta E_{eff}(y) = \Delta E_{SAW}(y) + \Delta E_{shift}(y)$ at a pump power 1.5 times above the threshold. At the wavevector of the driven mode $k = k_p$, the pump energy is detuned with respect to the polariton energy minima and maxima by 0.45 meV and 0.05 meV, respectively. The blue-shift due to the bistability induced transition nearly compensates the frequency detuning of the pump with respect to the unperturbed resonance. Accordingly, the maxima of the PL intensity correspond to the minima of the SAW potential. As a result, above the threshold the SAW potential influencing the condensate ($k = 0$) polariton mode appears screened by a factor of 3 with the resultant reduction of the energy gap from $\frac{V_{SAW}}{2} = 0.15$ meV down to 0.05 meV, which agrees well with the experiment.

The screening of the SAW potential is also revealed by the pump power dependence of the first diffraction maxima of the pump mode (Fig. 5b). Below the threshold the $|\psi_X|^2\psi_X$ terms are negligible with respect to the effective inter-mode polariton interaction caused by the inhomogeneity of the exciton landscape. If the latter is sinusoidal, the intensities of the diffraction replicas should increase proportionally to $V_{SAW}|\psi|^2$. Such a behaviour drastically changes on reaching the threshold: the jump in $|\psi|^2$ is accompanied by a sharp drop in the diffraction maxima since the potential is partially screened out and the diffraction rate lowered. Eventually at high pump powers the intensities of the pump replicas change very slowly in good qualitative agreement with the experiment.

**Incoherently pumped BEC condensates in SAW potential**

We now turn to the coherence properties of condensates formed under optical excitation at the exciton energy at a large angle of incidence of $30^0$ (indicated by $P_{BEC}$ in Fig. 1c). This excitation condition is very similar to that employed in Ref.[22] to excite a non-equilibrium polariton BEC. Figures 6a and b display dispersion images recorded at an optical power 1.5 to 2 times above the BEC condensation threshold $I_{BEC}(th) = 2.2 \times 10^4$ W/cm$^2$ in the absence and presence of a SAW (u = 3.3 mW$^{0.5}$), respectively. To achieve the high optical densities required for excitation in resonance with the exciton-like lower polariton branch, the experiments were carried out using an elliptical 8 × 25 µm excitation spot with the long axis aligned along the SAW direction. As in the OPO case, a macroscopically occupied s state forms at k=0 in the absence of the SAW. With the SAW, the relaxation from the high energy exciton-like states populates the minima of both the first and second folded MBZ, leading to condensation into the s and p states indicated on Fig. 1b. Such behavior, which has been reported for static gratings and explained by a rate equation model for the relaxation process, [15] is not observed for the OPO, where the phase matching conditions restrict condensation to the lowest folded branch. By contrast, the parametric process leading to population of the s′ zone boundary states found in the OPO cannot occur for the incoherent relaxation of the BEC case.

Figures 6c and d display the first-order spatial correlation function $g^{(1)}$(y,−y) of the s and p states, respectively. For the s states, the coherence length $L_s$ reduces from the unperturbed value of ~ 10 µm to ~ 5-7 µm under strong acoustic excitation (u = 3.3 mW$^{0.5}$). This clearly shows that the reduction of the spatial coherence of the s−states is directly related to the spatial confinement of the condensates. Surprisingly, the coherence length for the p states of $L_p$ ~ 13 µm (limited by the finite size of the laser spot) is greater than that found without a SAW. In contrast to the bonding s states, the p states have an anti-bonding character and lie energetically above the potential ($\Delta E_{SAW}$) [11].

The longer p state coherence is attributed to the larger extent of their wave function, which spreads over more than one SAW period. The fact that the coherence length of the p state (13 μm) exceeds the one for the unperturbed condensate (10 μm) may also be related to its symmetry which allows a greater spreading in real space.

The coexisting s- and p- type condensates, characteristic of the non-equilibrium polariton system,[31], interact with each other as well as with the modulated density of the pump and exciton reservoir. The simultaneous observation of the s− and p− condensates allows an estimation of the effective potential amplitude ($\Delta E_{eff}$) in the condensation regime. At low u the energy difference $\Delta E_{sp}$ between s− and p−condensates sets an upper limit for $\Delta E_{eff}$, which is slightly smaller than $\Delta E_{sp}$. By contrast, below condensation threshold at high u the flattening of the lowest dispersion branch leads to $\Delta E_{sp} \approx V_{SAW}/2$, as can be seen from Fig. 2d. By changing the rf-amplitude by a factor of 3 (from 1 mW$^{0.5}$ to 3.16 mW$^{0.5}$) the energy splitting $\Delta E_{sp}$ can be tuned from 0.1 to 0.16 meV, as shown in Fig. 7. A comparison with the case below threshold reveals that $\Delta E_{sp}$ is reduced above threshold by a factor of ~ 1.3 with respect to the energy gap $\Delta E_{gap} \approx V_{SAW}/2$ below threshold, as expected from partial screening of the SAW potential at high polariton densities.

We note that in the BEC case the spatial modulation of the pump field is strongly reduced compare to the OPO case due to the quite broad exciton linewidth (full width at half maximum ~ 0.6 meV), which is much larger than the spatial modulation of the exciton energy (0.2 meV at u=2.87 mW$^{0.5}$). In this case polariton-polariton interactions within the BEC s and p-states may play a more important role in the screening of SAW potential. This may explain why the gap between the 1$^{st}$ and the 2$^{nd}$ MBZ in the condensation regime is at least a factor of two larger than in the OPO case.

**Conclusion**

In conclusion we observe strong reduction of the energy gap in the excitation spectrum of OPO condensates due to strong modulation of the polariton density in anti-phase with the SAW potential. The observed results are explained quantitatively by numerical modelling of the OPO subject to the SAW potential using generalised Gross-Pitaevskii equations. In the case of incoherent pumping non-equilibrium coexisting condensates with s- and p- type wavefunctions are observed and the energy splitting between these condensate states is also reduced with respect to the gap of the one particle spectrum below threshold. However, the screening effects are less pronounced due to weaker modulation of the highly populated pump state.

**Acknowledgements**. We acknowledge support from EPSRC grants EP/G001642/1 and EP/J007544/1, EU ITN "Clermont 4", and RFBR grant 12-02-00799-a.

**Figures**

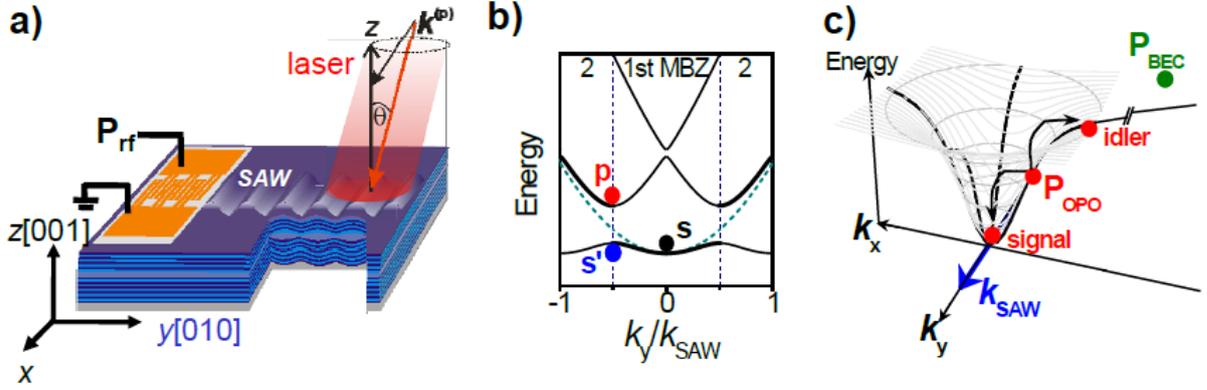

Fig.1 (Color online) a) Schematic of microcavity device with interdigitated transducer on top of the sample generating SAW along [110]. b) Calculated single particle polariton spectrum in periodic potential. c) Un-modulated polariton dispersion versus $k_x$ and $k_y$. Pump excitation conditions for optical parametric oscillation ($P_{OPO}$) and nonequilibrium polariton BEC ($P_{BEC}$) are shown.

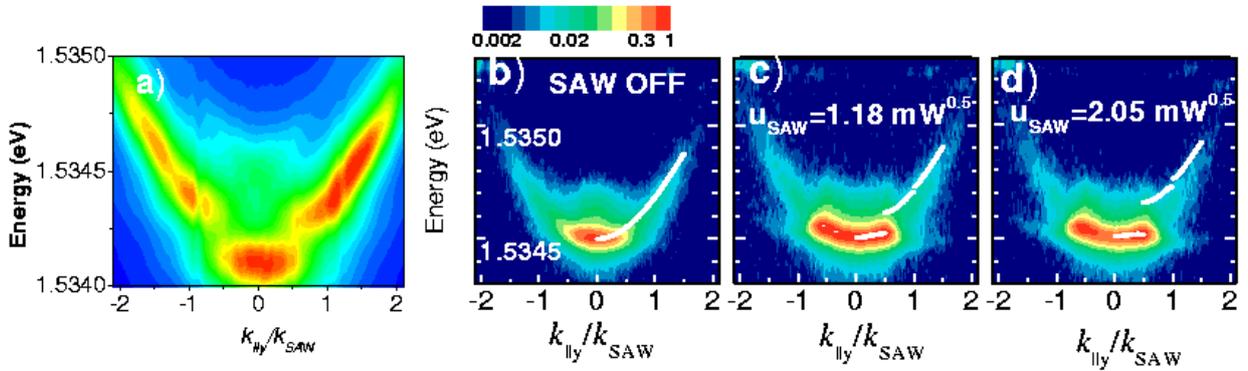

Fig.2 (Color online) Polariton emission spectra versus energy and $k_y$ taken at $k_x=0$ below (a) and above (b-d) condensation threshold of the OPO. Data in b) are measured without SAW. (c-d) Condensation with applied SAWs.

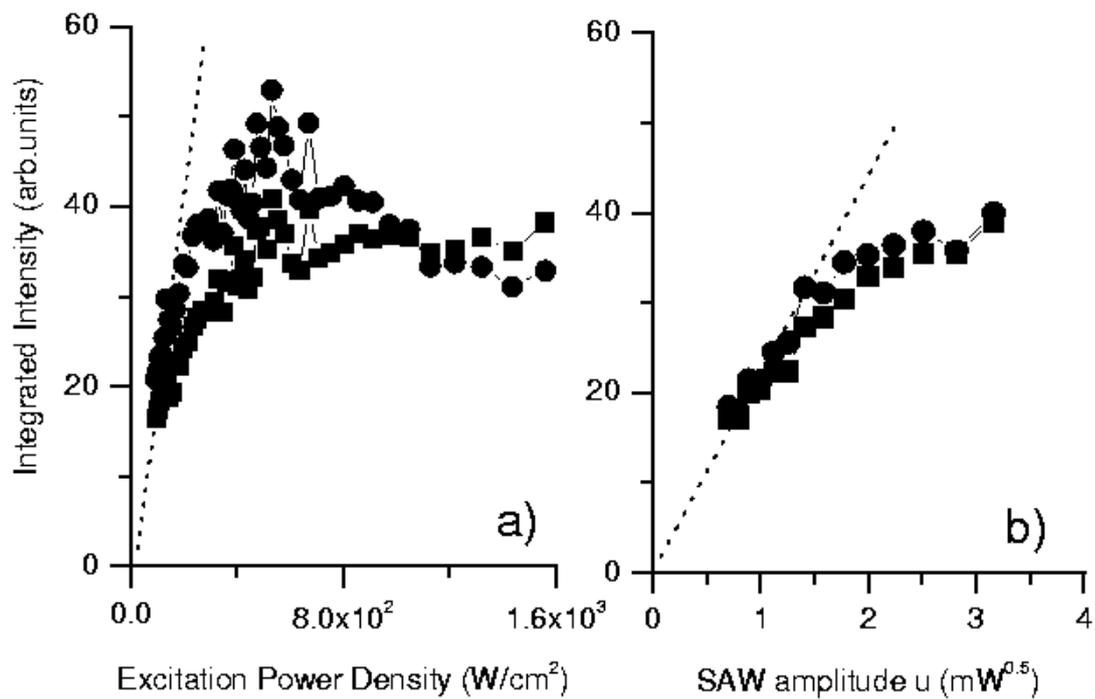

Fig.3 Intensities of first order diffraction replicas of the pump versus pump power (a) and SAW amplitude (b) at fixed u=2.8 mW$^{0.5}$ and $I_L$~800 W/cm$^2$, respectively.

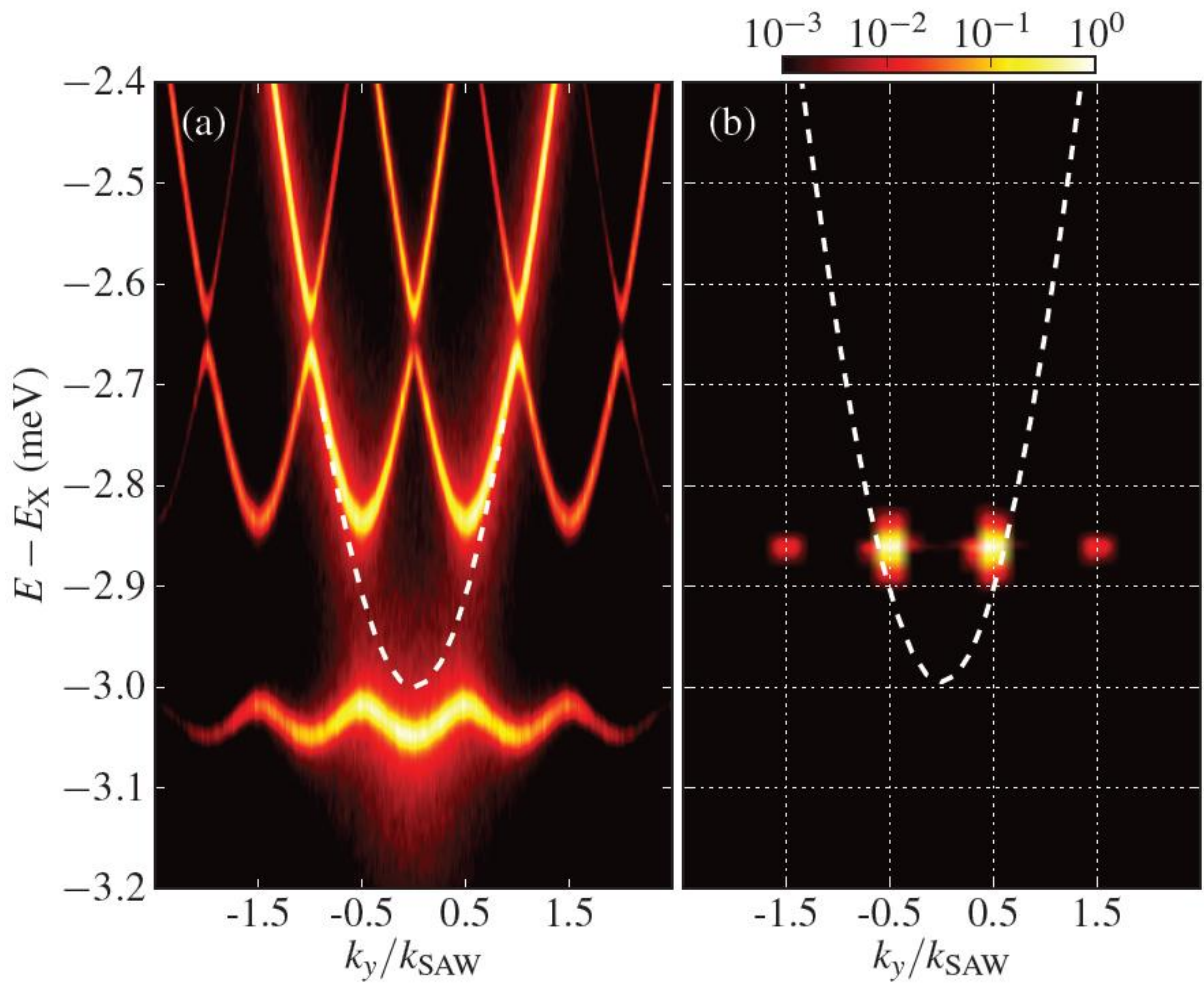

Fig. 4 (Color online) Calculated PL spectrum under a nearly zero pump power (a) and slightly above the threshold of parametric generation (b).

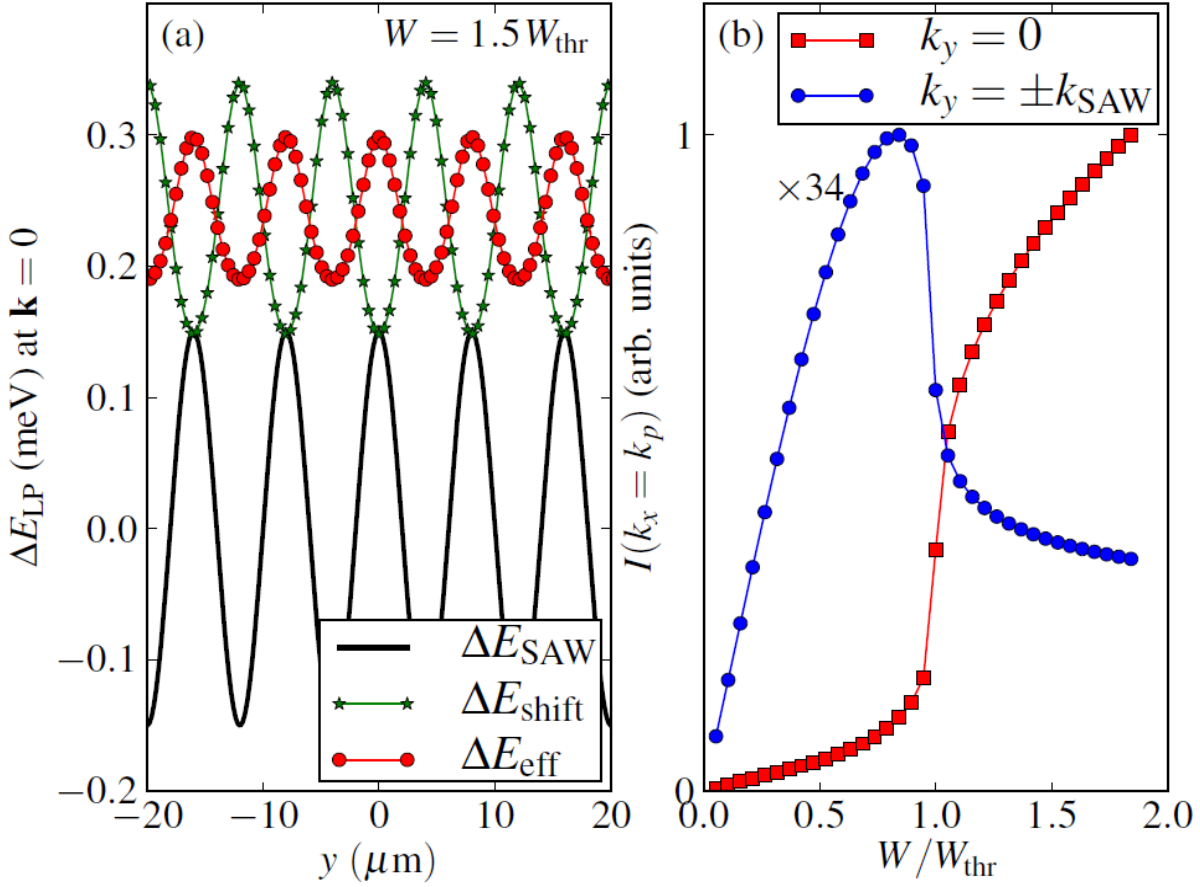

Fig. 5 (Color online) (a) Model dependences of the SAW-induced LP potential ($\Delta E_{\mathrm{SAW}}$), LP blue-shift induced by the polariton-polariton interaction at a pump power 1.5 times above the threshold ($\Delta E_{\mathrm{shift}}$), and the effective polariton energy, the sum of the above two contributions ($\Delta E_{\mathrm{eff}}$). (b) Intensities of the driven mode ($k_y = 0$) and diffracted pump replicas ($k_y = \pm k_{\mathrm{SAW}}$) vs. pump power expressed in units of its threshold magnitude.

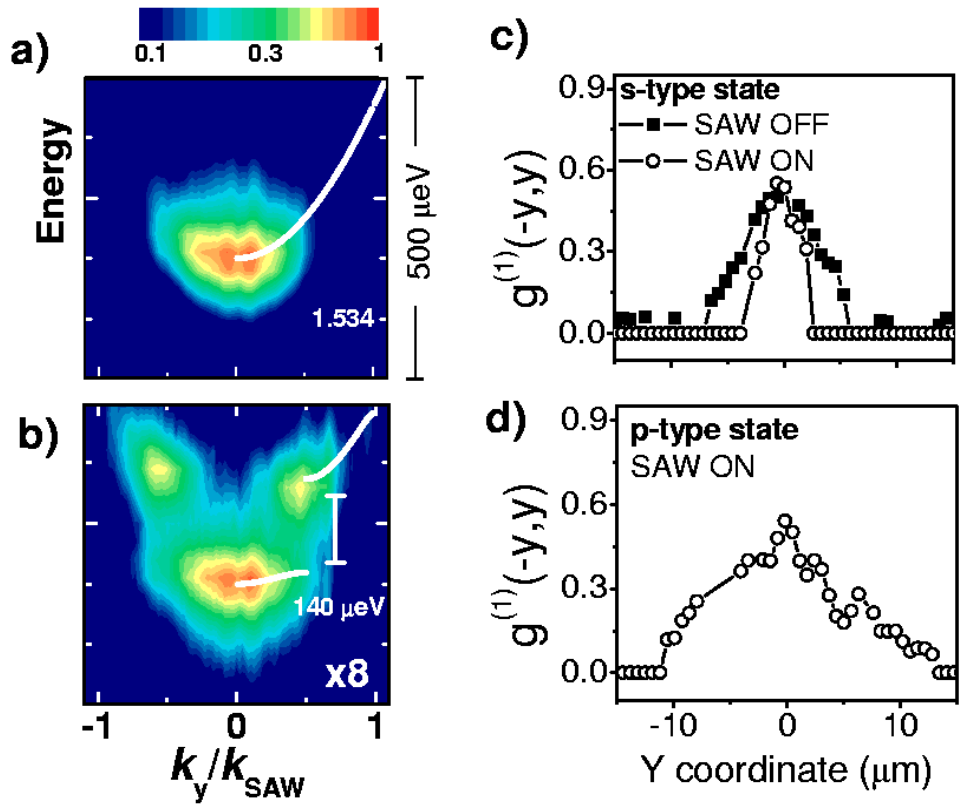

Fig.6 (Color online) Polariton emission versus energy and $k_y$ recorded at an optical power 1.5 to 2 times above the BEC condensation threshold $I_{BECth}$ = 2.2 × $10^4$ W/cm$^2$ in the absence (a) and presence (b) of a SAW (u = 3.3 mW$^{0.5}$), respectively. The first-order spatial correlation function $g^{(1)}$(y,−y) of the s and p states, respectively is shown in (c) and (d)

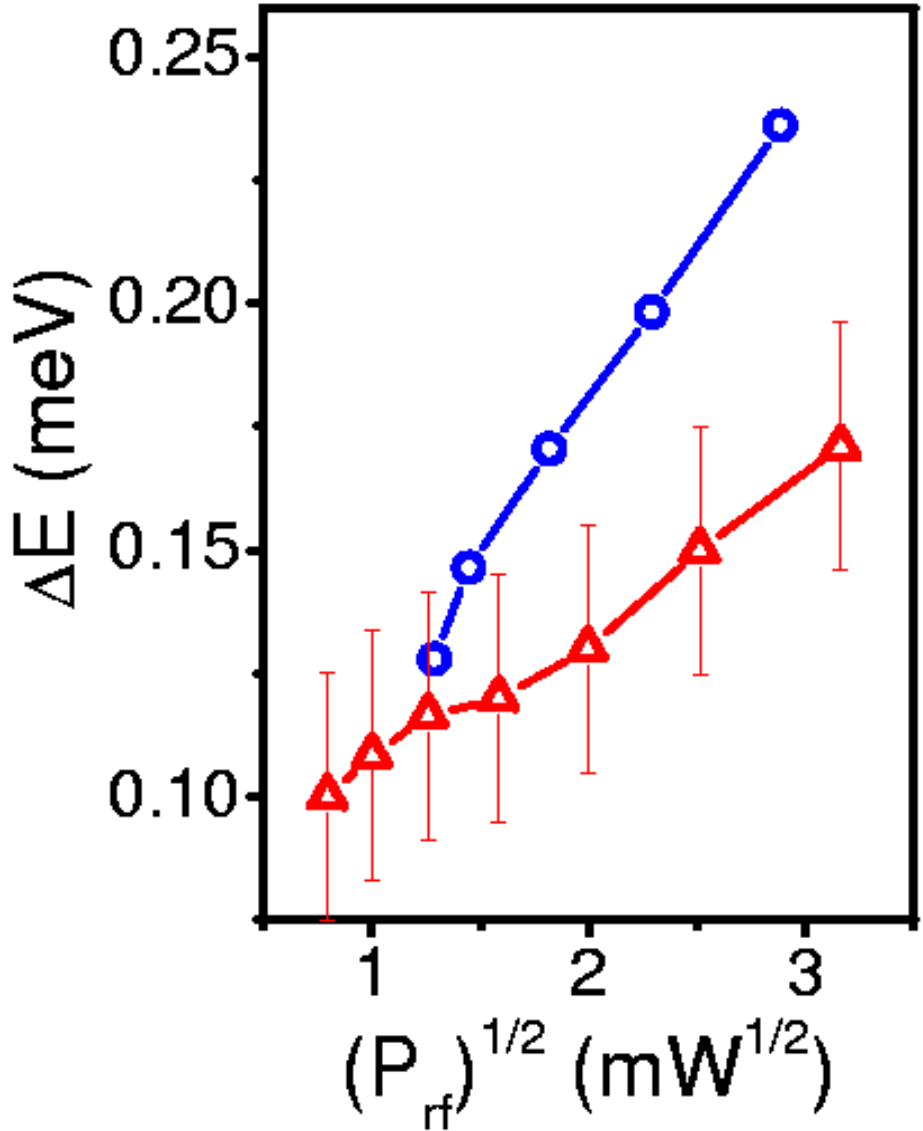

Fig.7 (Color online) Energy splitting between s and p states of BEC condensate versus SAW amplitude is shown by red triangles. Polariton energy gap at the edge of the 1st MBZ at $k_{SAW}/2$ below condensation threshold is shown by blue circles.